\documentclass[fleqn,usenatbib]{mnras}

\usepackage{newtxtext,newtxmath}

\usepackage[T1]{fontenc}
\usepackage{siunitx}
\usepackage{upgreek}
\usepackage{subcaption}
\usepackage{xcolor}

\DeclareRobustCommand{\VAN}[3]{#2}
\let\VANthebibliography\thebibliography
\def\thebibliography{\DeclareRobustCommand{\VAN}[3]{##3}\VANthebibliography}

\usepackage{graphicx}	
\usepackage{amsmath}	

\title[Scattering Transparency Effect]{Scattering Transparency of Clouds in Exoplanet Transit Spectra}

\author[B. Jaiswal and T. D. Robinson]{
Bhavesh Jaiswal$^{1,2}$\thanks{E-mail: bhavesh@ursc.gov.in}
and Tyler D Robinson$^{3,4,5,6}$
\\
$^{1}$Space Astronomy Group, U R Rao Satellite Centre, Bengaluru 560037, India\\
$^{2}$Dept. of Physics, Indian Institute of Science, Bengaluru 560012, India\\
$^{3}$Department of Planetary Sciences; Lunar and Planetary Laboratory, University of Arizona, Tucson, AZ 85721, USA\\
$^{4}$Department of Astronomy and Planetary Science, Northern Arizona University, Flagstaff, AZ 86011, USA\\
$^{5}$Habitability, Atmospheres, and Biosignatures Laboratory, University of Arizona, Tucson, AZ 85721, USA\\
$^{6}$NASA Nexus for Exoplanet System Science Virtual Planetary Laboratory, University of Washington, Box 351580, Seattle, WA 98195, USA
}

\date{Accepted XXX. Received YYY; in original form ZZZ}

\pubyear{2023}

\begin{document}
\label{firstpage}
\pagerange{\pageref{firstpage}--\pageref{lastpage}}
\maketitle

\begin{abstract}
The presence of aerosols in an exoplanet atmosphere can veil the underlying material and can lead to a flat transmission spectrum during primary transit observations. In this work, we explore forward scattering effects from super-micron sized aerosol particles present in the atmosphere of a transiting exoplanet. We find that the impacts of forward scattering from larger aerosols can significantly impact exoplanet transits and the strength of these effects can be dependent on wavelength. In certain cloud configurations, the forward-scattered light can effectively pass through the clouds unhindered, thus rendering the clouds transparent. The dependence of the aerosol scattering properties on wavelength can then lead to a positive slope in the transit spectrum. These slopes are characteristically different from both Rayleigh and aerosol absorption slopes. As examples, we demonstrate scattering effects for both a rocky world and a hot Jupiter. In these models, the predicted spectral slopes due to forward scattering effects can manifest in the transit spectrum at the level of $\sim$10s to $\sim$100s of parts per million and, hence, could be observable with NASA's \textit{James Webb Space Telescope} .

\end{abstract}




\section{Introduction}
Transiting exoplanets provide an opportunity to study the atmospheric structure of these distant worlds through transit spectroscopy. The small amount of stellar light which passes through the narrow annulus of the exoplanet atmosphere in the transit geometry interacts with the atmosphere and thus can carry information about the atmospheric physical and chemical state \citep[e.g.][]{2000ApJ...537..916S,2001ApJ...553.1006B,2001ApJ...560..413H}. Spectroscopic signatures of the atmosphere for a transiting exoplanet are subtle, in part due to the small size of planets as compared to stars and due to the relatively small covering area of the atmospheric annulus as compared to the planetary disk. In the last two decades, exoplanet transit spectroscopy has moved from noisy detections of simple atmospheric constituents \citep[e.g.][]{2002ApJ...568..377C,2007Natur.448..169T,2008NatureSwain} to, more recently, searches for trends \citep[e.g.][]{2015PASP..127..941C, 2016Natur.529...59S} in the transit spectra of distinct types of worlds. Promisingly, early results from \textit{James Webb Space Telescope} (\textit{JWST}) already show detections of atmospheric species from transit spectra \citep{2023Natur.614..649J,2023Natur.614..664A,2023Natur.614..659R}.

\begin{figure*}
$\begin{array}{rl}
    \multicolumn{2}{c}{\includegraphics[width=0.9\textwidth]{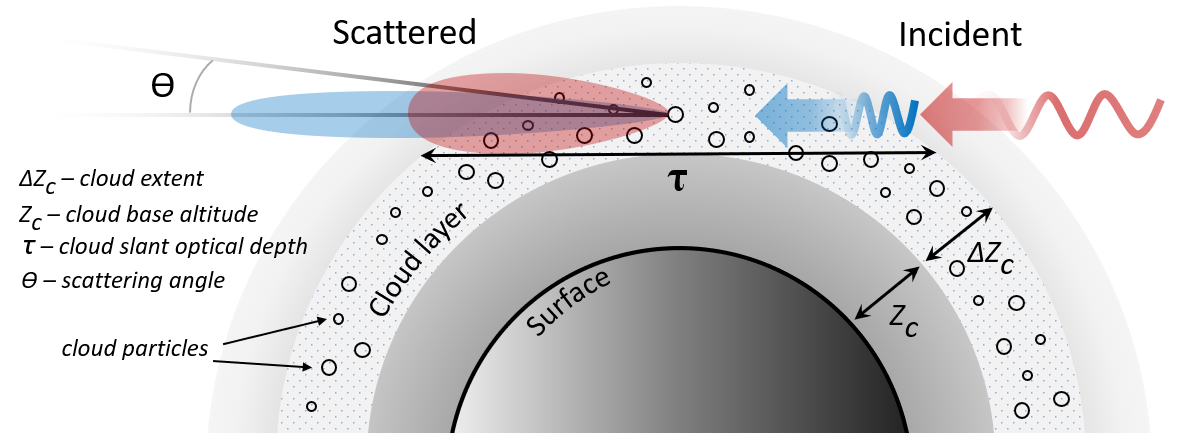}}\\\\\\\\
    \includegraphics[width=0.5\textwidth]{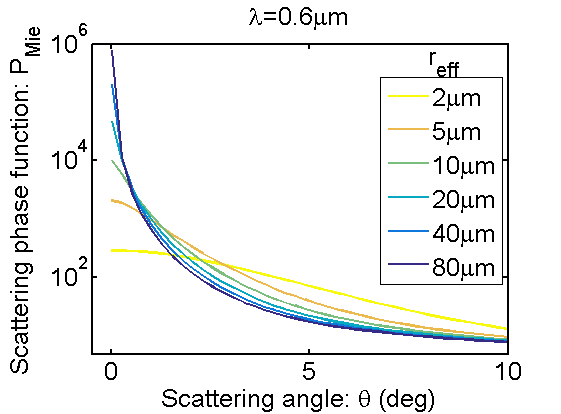}&
    \includegraphics[width=0.5\textwidth]{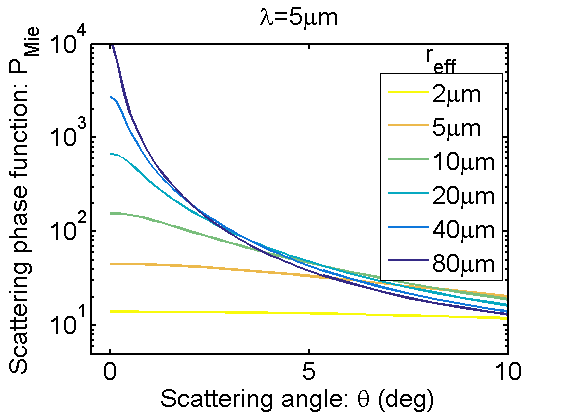}\\
\end{array}$
\caption{[Top] Cloud scattering in an exoplanet atmosphere. Strong forward scattering from the cloud particles is shown for the slant incident direction. Shorter wavelengths are more-strongly scattered in the forward direction than longer wavelengths.[Bottom] The Mie scattering phase function for various sizes ($r_{\mathrm{eff}}$, in legends) of water cloud particles shown at two wavelengths: $\lambda$=0.6\,$\upmu$m  (left) and $\lambda$=5\,$\upmu$m (right)}\label{fig:fig1}
\end{figure*}

\begin{figure*}
$\begin{array}{rl}
    \includegraphics[width=0.5\textwidth]{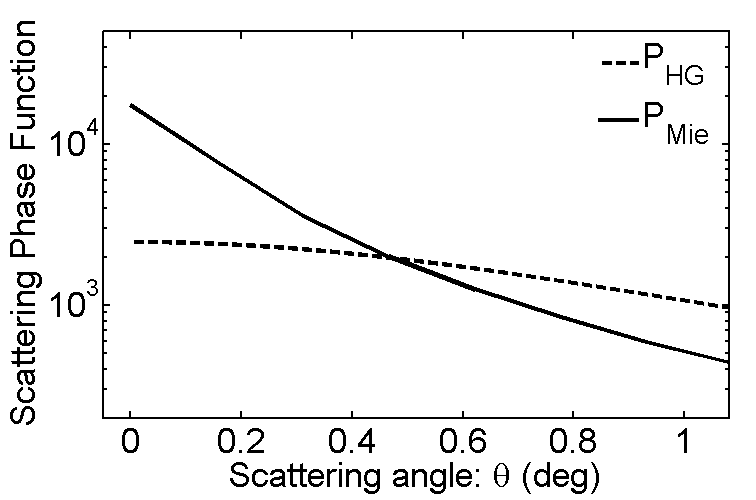}&
    \includegraphics[width=0.5\textwidth]{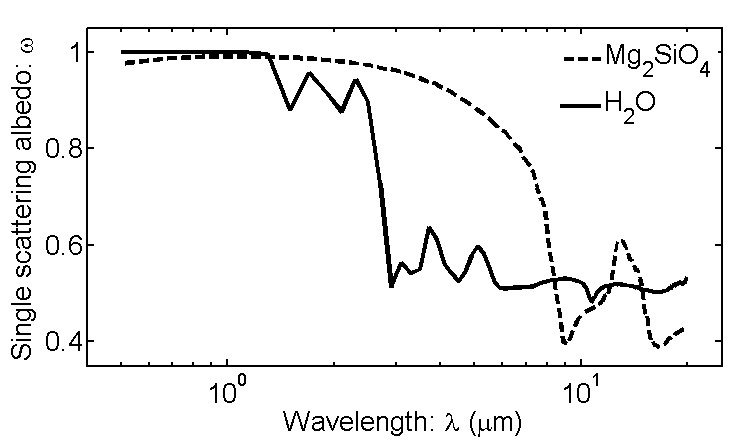}\\
\end{array}$
\caption{[Left] Scattering phase function obtained from Mie theory and its approximation by the Henyey-Greenstein function. $P_{\mathrm{Mie}}$ is calculated at a wavelength of 0.6\,$\upmu$m for water ice particle with $r_{\mathrm{eff}}$ of 20\,$\upmu$m. $P_{\mathrm{HG}}$ is calculated using Eq. \ref{eq:PP} for $\theta_s$ of $\ang{1}$. [Right] Single scattering albedo ($\omega$) for H${_2}$O ice and Mg${_2}$SiO${_4}$ cloud particles.  \label{fig:fig2}}
\end{figure*}

Unfortunately, only a minority of the thousands of currently-known exoplanets are suitable for atmospheric reconnaissance. In many of these cases, existing observations indicate the presence of clouds or hazes \citep{2008MNRAS.385..109P,2009A&A...505..891S,2010Natur.468..669B,2014ApJ...794..155K,2019NatAs...3..813B, 2019RNAAS...3....7W,2023Natur.614..649J}. Generally, these aerosols have presented themselves either directly as slopes of decreasing transit depth with increasing wavelength (potentially characteristic of sub-micron particles \citep[e.g.][]{2008A&A...481L..83L}) or somewhat indirectly through a continuum absorption that suppresses gas absorption features \citep[e.g.][]{2014Natur.505...69K}. There have been limited observations of wavelength dependent extinction \citep{2019NatAs...3..813B} which can also arise due to clouds. The composition of exoplanet clouds and hazes largely remains a mystery as any characteristic spectral signatures are either weak or outside the spectral range of observations with the pre-\textit{JWST} facilities, albeit the early broadband \textit{JWST} results are encouraging \citep{2022arXiv220900620M}. Nevertheless, studies have predicted the kinds of spectral slopes or absorption features that could be expected for hazes in hot Jupiter atmospheres \citep{2015A&A...573A.122W,10.1093/mnras/stx1849,2020ApJ...895L..47O}. We find that the studies which have investigated the forward scattering from clouds so far \citep[see for e.g.][]{2020ApJ...889..181S} have almost always considered sub-micron sized scatterers which have a characteristic negative spectral slope resembling Rayleigh-like scattering. The modeling investigations, on the other hand, spanning hot Jupiters to sub-Neptunes have shown the potential for higher-altitude clouds and hazes at sizes extending beyond 1\,$\upmu$m and maybe even beyond 10$\upmu$m \citep{2018ApJ...860...18P,2021A&A...646A.171C}. Additionally, the clouds on temperate rocky worlds can also be expected to grow much beyond 1\,$\upmu$m in resemblance to clouds on Earth \citep[see for e.g.][]{2001JGR...10627449D}. In such a scenarios, it is imminent to explore the spectral effect of forward scattering from super-micron sized aerosols.

 Unlike smaller (i.e.,~sub-micron haze) particles, the super-micron sized aerosol particles can have very strong forward scattering at wavelengths where the size parameter of particles ($2\pi r_{\mathrm{eff}}/\lambda$) is much larger than unity. This scattering behaviour is typical of Mie scattering. Depending upon the clouds and the transit configuration, this forward scattering can be so strong that deeper atmospheric layers within the cloud can be probed, as compared to a more-isotropically scattering aerosol scenario \citep{2012Icar..221..517D,Robinson_2017,2018MNRAS.473.1801G}. Thus, the upper reaches of some clouds may appear transparent at shorter wavelengths but transition to opaque conditions at longer wavelengths as the size parameter changes from much larger than unity to of order (or smaller than) unity, as predicted in \citet{Robinson_2017}. This variation in transparency can occur over a large spectral range, making the effect more-straightforward to detect.

In what follows, we investigate the spectral scattering effects of aerosols for transiting exoplanets. For the majority of the simulations considered here we omit gas absorption to focus on the impacts on the continuum set by the cloud/haze. This work doesn't seek to delineate between clouds and hazes and explores a range of aerosol profiles. For convenience, though, we use the term "clouds" throughout. In Section~\ref{sec:forw_scat} of this paper we present the theory of forward scattering for large cloud particles in the transit geometry. Section~\ref{sec:RT_calc} discusses a Monte Carlo radiative transfer model which is used here for more-detailed simulations. In Section~\ref{sec:Res} we present the results of simulations for prescribed cloud structures in model atmospheres for a temperate planet TRAPPIST-1e and a hot Jupiter HD\,189733b. Finally, we briefly discuss the feasibility and limitations of the simulations in section~\ref{sec:Disc} and conclude in section~\ref{sec:Concl}.

\section{Forward Scattering Theory} \label{sec:forw_scat}
Mie scattering from particles whose size parameter is larger than the wavelength has a characteristic scattering phase function where a significant amount of light is scattered in the forward direction \citep{1974SSRv...16..527H}. The scattering phase function is extremely large at \ang{0} scattering angle (i.e.,~the forward-scattering direction) and decreases rapidly for larger angles. Figure~\ref{fig:fig1} shows the Mie scattering-derived phase function of water droplets of various sizes at wavelengths of 0.6\,$\upmu m$ and 5\,$\upmu m$ where the sharp rise of the scattering phase function at forward scattering angles is clearly seen for shorter wavelengths. Importantly, the forward-scattering angle is probed in the exoplanet transit geometry where stellar photons can be scattered in the same direction of travel from the limb of the exoplanet towards the direction of the observer \citep{2012Icar..221..517D}. As is apparent in Figure~\ref{fig:fig1}, light at shorter wavelengths is forward-scattered more effectively than at the longer wavelengths. The angular extent of the star apparent from the planet is an important parameter here as the incident light is sampled from all the parts of the stellar disc. The forward scattering phase function essentially gets convolved over the angular size of the star ($\sim\ang{7}$ for HD\,189733b and $\sim\ang{1}$ for TRAPPIST-1e system).

\begin{figure*}
$\begin{array}{rl}
    \includegraphics[width=0.5\textwidth]{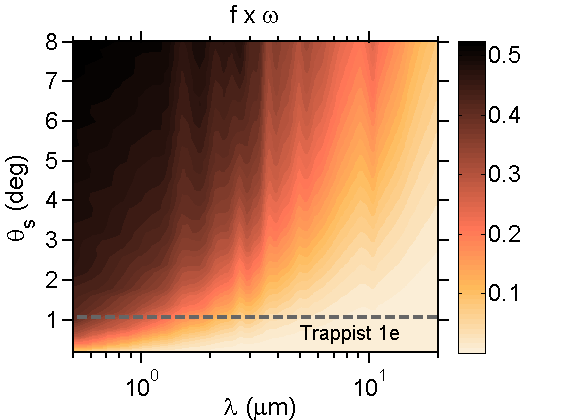}&
    \includegraphics[width=0.5\textwidth]{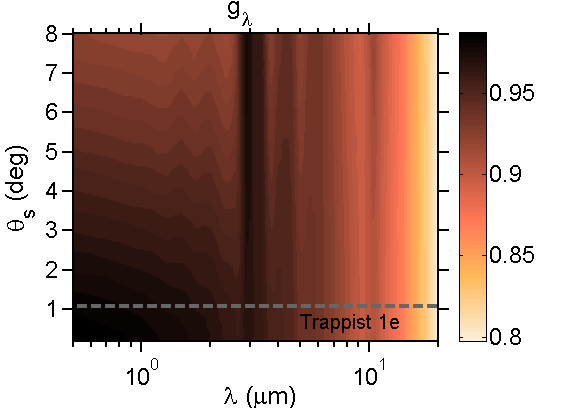}\\
    \includegraphics[width=0.5\textwidth]{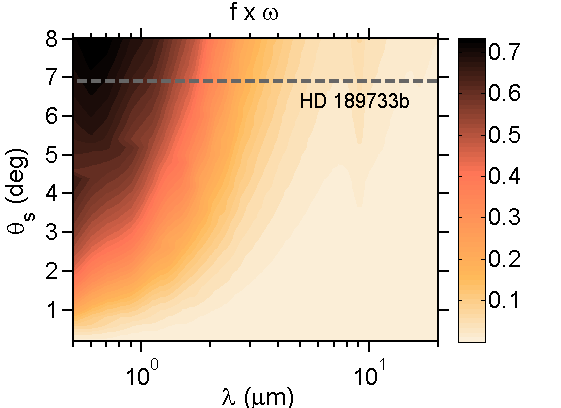}&
    \includegraphics[width=0.5\textwidth]{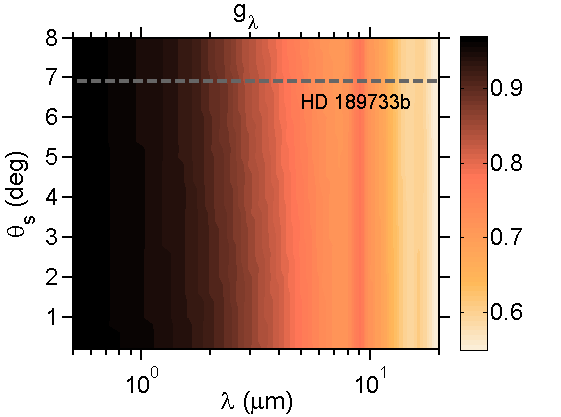}\\
\end{array}$
\caption{Spectral variations of $f\times\omega$ and $g$ for various $\theta_{\mathrm{s}}$. [Top] For TRAPPIST-1e with water ice particles with $r_{\mathrm{eff}}$=20\,$\upmu$m. [Bottom] For HD\,189733b with silicate particles with $r_{\mathrm{eff}}$=5\,$\upmu$m. Also marked are the location of $\theta_{s}$ corresponding to TRAPPIST-1e and HD\,189733b.  \label{fig:fig3}}
\end{figure*}

For applications in planetary atmospheres, the shape of the scattering phase function ($P(\theta)$; where $\theta$ is the scattering angle) is often approximated with a Henyey-Greenstein (H-G) scattering phase function, whose analytic nature makes it useful for radiative transfer calculations. The H-G phase function \citep[see e.g.][for its application to exoplanets]{ refId0,2015ApJ...811...18M,Robinson_2017} is characterised by a single parameter\,---\,$g$, which controls the shape of the phase function and the strength of the forward peak\,---\,and is given by,
\begin{equation}
P_{\rm HG}(\theta,g) = \frac{1}{4\pi} \frac{1-g^2}{\left( 1+g^2-2g\cos\theta \right)^{3/2}} \ . \label{eq:P}
\end{equation}

It is worth noting here that both, the HG parameter $g$ and the phase function $P_{HG}$ are wavelength specific. Following \cite{2017ApJ...850..128R}), aerosol forward scattering can be incorporated into the transmission of stellar intensity along the limb of a transiting exoplanet by writing the equation of transfer as,
\begin{equation}
    \frac{dI}{d\tau_{\mathrm{eff}}} = -I \ , \label{eq:dI}
\end{equation}
 where I is the radiant intensity along the path and $d\tau_{\mathrm{eff}}$ is the differential ``effective'' aerosol optical depth along the slant path. Now, we define a correction factor that is calculated by integrating the scattering phase function in the forward direction,
\begin{equation}
f = 2\pi\int_{0}^{\theta_{\mathrm{S}}} P(\theta) \sin(\theta) \,d(\theta).\label{eq:f}
\end{equation}
  This correction factor is dictated by the angular extent of the star ($\theta_{\mathrm{S}}$) as seen from the planet \citep{2018MNRAS.473.1801G}. 
 Considering this correction factor, effective optical depth is designed to account for a lossless process where slant-path light that interacts with the medium is simply scattered into the forward direction, with, 
\begin{equation}
    d\tau_{\mathrm{eff}} = (1-f\omega)d\tau_{\mathrm{ext}} \ , \label{eq:dtau}
\end{equation}
where $\omega$ is the single scattering albedo. 
Note that Equation~\ref{eq:dI} ignores the multiple-scattering source term which comes from the isotropic component of the phase function and is negligible for transiting geometry \citep[explained in][]{2017ApJ...850..128R}. As seen in Figure~\ref{fig:fig1}, the scattering phase function can vary drastically with wavelength, thereby leading to variations in both $f$ and $d\tau_{\mathrm{eff}}$ across the spectral range. Shorter wavelengths, due to a steeper rise of the phase function, can have a significantly smaller $d\tau_{\mathrm{eff}}$ causing a transparency in the clouds not encountered at longer wavelengths. The wavelength-dependent factor $f\times\omega$ is a measure of this transparency effect.

Alongside the more-intuitive analytic theory described above, we explore more-detailed scattering solutions using a Monte Carlo-based approach. We adopt the publicly-available \texttt{scaTran} package \citep[discussed in][]{Robinson_2017}, which includes scattering, absorption, and refraction effects for exoplanet atmospheres in the transiting geometry. Mie scattering by aerosol particles is considered using the Monte Carlo (MC) algorithm by sampling the H-G phase function, where the wavelength-dependent H-G asymmetry parameter is designed to reproduce the forward scattering peak from the Mie-derived phase function. Thus, at each wavelength we determine the value of $g$ that satisfies, 
 \begin{equation}
\int_{0}^{\theta_{S}} P_{\rm Mie}(\theta) \sin\theta \,d\theta = \int_{0}^{\theta_{S}} P_{\rm HG}(\theta,g) \sin\theta \,d\theta \ , \label{eq:PP}
\end{equation}
where $P_{\mathrm{Mie}}$ is calculated using Mie theory. Figure \ref{fig:fig2} shows $P_{\mathrm{Mie}}$ and $P_{\mathrm{HG}}$ for a water cloud particle (with $r_{\mathrm{eff}}$ = 20\,$\upmu$m) which satisfies Equation~\ref{eq:PP}. It is noteworthy that our choice of 'g' parameter depends upon integrating the phase function for the stellar angular size ($\theta_{S}$) with an intention to preserve the total scattering in the direction of the star. As is seen in the figure, this HG phase function underestimates the forward scattering by an order or magnitude at $\sim\ang{0}$ but it overestimates it proportionally at $\sim\ang{1}$ and hence maintaining the total scattering in the direction of the star. Here, a choice of a multi-parameter HG phase function might be more useful for a better fit to the forward part of the Mie scattering phase function. However it will add more parameters (and complexity) in our model and hence we restrict the present analysis to a single parameter HG function. The single scattering albedo, $\omega$, is also  shown in Figure~\ref{fig:fig2} for the water and silicate clouds. The optical constants for the scattering calculations of the clouds are considered from \cite{2015MNRAS.454....2B}.

\section{Radiative Transfer Simulations}\label{sec:RT_calc}
We prescribe a simple cloud model for full-scale MC simulations, as is shown in Figure~\ref{fig:fig1}. This model contains a layer of cloud particles spread uniformly around the terminator. In the transit geometry, we explore the effect of clouds mainly with three parameters: cloud base height ($Z_{\mathrm{c}}$), cloud vertical extent ($\Delta{Z}_{\mathrm{c}}$) and cloud slant scattering optical depth ($\tau$, at $\lambda$=0.6\,$\upmu$m), which are all depicted in Figure~\ref{fig:fig1}. We study two cases of planets: a rocky world - TRAPPIST-1e \citep{2017Natur.542..456G}, with water ice clouds and a hot Jupiter - HD\,189733b \citep{2005A&A...444L..15B}, with silicate clouds. The clouds can exist in a variety of conditions on temperate as well as hot jupiter planets. For simplicity, we consider uniformly distributed clouds having a log-normal size distribution which is given as 

\begin{equation}
n(r) = \frac{1}{r} \exp \left(-0.5\frac{(\ln{r}-\ln{r_g})^2}{(\ln{\sigma})^2}\right) \ . \label{eq:nr}
\end{equation}
 Here, $n(r)$ is the number density of cloud particles with radius $r$ where the modal radius ($r_g$) and width of the distribution ($\sigma$) are related to effective radius $r_{\mathrm{eff}}$ and effective variance $v_{\mathrm{eff}}$ as:
 
\begin{equation}
r_{\mathrm{eff}} =  \frac{r_g}{\exp(-2.5(\ln{\sigma})^2)} \ , \label{eq:reff}
\end{equation}

\begin{equation}
v_{\mathrm{eff}} =  \exp((\ln{\sigma})^2)-1\ . \label{eq:veff}
\end{equation}

The cloud vertical optical depth ($\tau$) is considered as decreasing exponentially above the cloud base ($Z_{c}$) i.e. for $z>Z_{c}$ 

\begin{equation}
\tau(z) =  \tau(Z_{c})\exp{\left(-\frac{z}{Z_H}\right)}\ , \label{eq:tau}
\end{equation}
where $Z_{\mathrm{H}}$ is the scale height of the cloud. Below the cloud base the cloud optical depth is zero.

In order to demonstrate the effect, a fiducial value of $r_{\mathrm{eff}}$ of 5\,$\upmu$m \citep[similar to silicate clouds of][]{2018ApJ...860...18P} and 20\,$\upmu$m \citep[similar to water ice clouds of][]{article} is used for HD\,189733b silicate clouds and TRAPPIST-1e water ice clouds respectively. The key model parameters adopted in the simulations are given in Table~\ref{tab_model}.

Next, the evolution of spectral slope is discussed using two approaches -- the analytical approach (eq. \ref{eq:dI}-\ref{eq:f}) and a full scale 3D Monte Carlo (MC) scattering approach of \texttt{scaTran}. It is noteworthy that the analytical approach ignores the multiple scattering of photons whereas the MC approach does not. However, the effect of multiple scattering term is negligible in the analytical approach and, as we show later, the simulation results of both the approaches lead to a positive slope in the spectrum.

\begin{figure*}
\includegraphics[width=0.8\textwidth]{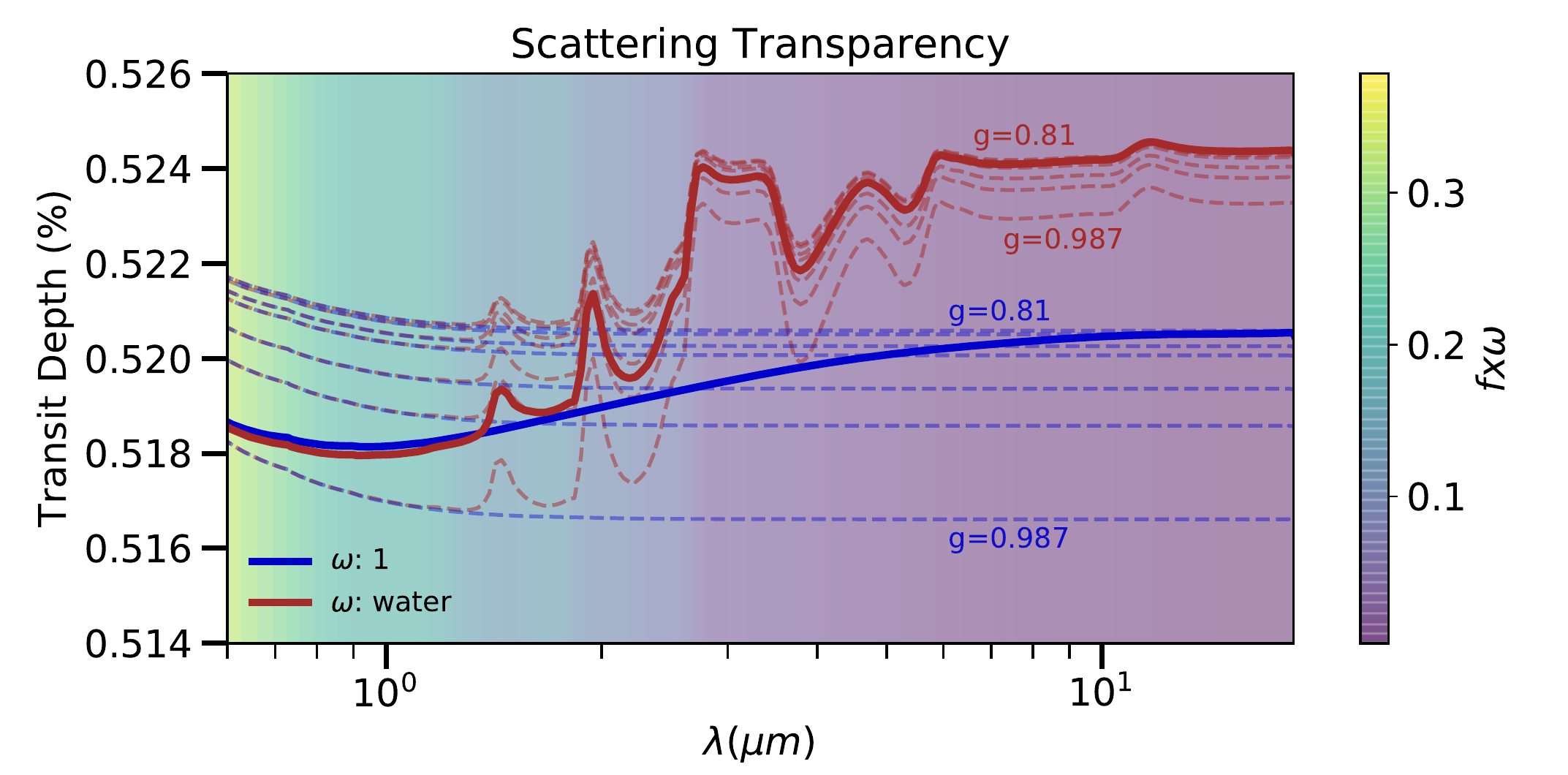}

\caption{Manifestation of the scattering transparency effect in the transit spectrum of TRAPPIST-1e system having a water cloud with $\tau$=1 and $Z_{\mathrm{c}}$=100 km. The transit spectrum is simulated with MC calculations for cloud particles having an $r_{\mathrm{eff}}$=20\,$\upmu$m. The dashed lines are the spectra obtained for constant values of $g$. The actual spectra, after considering the spectral variations of $g$, are shown in bold lines. They are obtained by interpolating among the dashed spectra. The blue curves do not consider the spectral variations of $\omega$ whereas the red curves consider the realistic variation $\omega$ for water clouds which leads to the spectral features of water droplets. Gas absorption is not considered. The spectral rise towards shorter wavelengths in the dashed curves is due to Rayleigh scattering. A color gradient for $f\times\omega$, which is a measure of scattering transparency, is shown in the background. This gradient of $f\times\omega$ is same as that shown in figure \ref{fig:fig3} for TRAPPIST-1e system. \label{fig:fig4}}
\end{figure*}

\subsection{The Analytical approach}
The spectral variation of the theoretically calculated parameters $f\times\omega$ and $g$ is shown in Figure~\ref{fig:fig3}. Results are shown for the water ice cloud and silicate cloud particle for a range of $\theta_{s}$ spanning that of TRAPPIST-1e to HD\,189733b. The variation of $f\times\omega$ and $g$ with wavelength as well as angular size ($\theta_{s}$) gives an intuitive understanding of the results via eq. \ref{eq:P}-\ref{eq:f}. Significantly larger values of $f\times\omega$ and $g$ at shorter wavelengths indicate the importance of a wavelength-dependent scattering transparency effect. It arises due to the sharp increase of the size parameter ($2\pi r_{\mathrm{eff}}/\lambda$) towards short wavelengths which leads to a strong forward scattering phase function.  Faintly visible vertical stripes in the $f\times\omega$ plot are a result of the spectral variation in $\omega$, as seen in Figure~\ref{fig:fig2}. The vertical stripes in the $g$ plots are a result of the variation in the real and imaginary part of the refractive index of the cloud particles. The variation of $f\times\omega$ with wavelength is essentially the variation in the effective optical depth $d\tau_{\mathrm{eff}}$ (via eq. \ref{eq:dtau}) which manifests as a spectral slope in the transit spectrum.

\subsection{The Monte Carlo approach}\label{sec:MC}
 After creating the model, the values of $g$ and $\omega$ are calculated for the given $r_{\mathrm{eff}}$ used in the model for all the wavelengths (as shown in Figure \ref{fig:fig3}) . The parameters $\tau$, $Z_{\mathrm{c}}$ and $Z_{\mathrm{H}}$ are left free for exploration. The cloud vertical extent ($\Delta{Z}_{\mathrm{c}}$) is always fixed at 4$\times Z_{\mathrm{H}}$.  Multiple instances of the \texttt{scaTran} model are run, each time for a fixed value of $g$. Each instance takes about $\sim10$ minutes to run on a quad-core machine. Then, to produce a spectrum having a variation in $g$ across wavelengths, we interpolate on these spectra. For TRAPPIST-1e, the values of $g$ varies from 0.987 (at $\lambda$=0.5\,$\upmu$m) to 0.81 (at $\lambda$=20\,$\upmu$m) for $r_{\mathrm{eff}}$ of 20\,$\upmu$m and for HD\,189733b this variation is from 0.98 (at $\lambda$=0.5\,$\upmu$m) to 0.55 (at $\lambda$=20\,$\upmu$m) for $r_{\mathrm{eff}}$ of 5\,$\upmu$m. 

Figure~\ref{fig:fig4} shows an example of how the wavelength-dependent scattering transparency is evolved from the interpolating the spectra, for the case of TRAPPIST-1e. The spectra obtained with various constant values of $g$ (as would be done without considering the scattering transparency effect) are shown with dashed lines. Considering the variation of $g$ with wavelength, the interpolated spectrum is shown with bold lines. As the transit depth is much lower at shorter wavelengths than at longer wavelengths the spectrum shows a positive slope in wavelength. This spectral shape closely resembles the variation of $f\times\omega$ with wavelength (shown as a color gradient in the background). To demonstrate the case of a scattering cloud without any spectral variation in scattering albedo ($\omega$), we show another spectrum (in blue) which has $\omega$ set to a fixed value of unity across the spectral range. Here too, we see the similar variation of transit depth in the spectrum (i.e. smaller transit depths at shorter wavelengths). In general, it can be said that the clouds can be more transparent at shorter wavelengths than at longer wavelengths if the scattering particles are sufficiently large with respect to the wavelength.

\begin{table}
	\caption{Key Model Parameters}\label{tab_model}
	\begin{center}
	\begin{tabular}{ l  l  l   }
		\hline
		                                 & \textbf{TRAPPIST-1e}    & \textbf{HD\,189733b}    \\\hline 
		 Model base pressure (Pa)        & $10^{5}$                 &  $10^{6}$          \\\hline   
		 Model top pressure (Pa)         & $10^{-5}$                &  $10^{-4}$       \\\hline 
		 Isothermal Temperature (K)      & 275                      &  1500                       \\\hline 
		 Atmospheric Composition         & N$_{2}$                  & H$_{2}$ \& He (0.85:0.15) \\\hline
		 Cloud                           & H${_2}$O                 & Mg${_2}$SiO${_4}$ (Forsterite)               \\\hline
		 Cloud $r_{\mathrm{eff}}(v_{\mathrm{eff}})$        & $20 \upmu m (2)$             & $5 \upmu m (0.5)$            \\\hline
      Layers                          & 180                     & 126       \\\hline
		 Gravity (m\,s$^{-2}$)           & 8.9                      & 21.8          \\\hline
		 Star radius (R\textsubscript{\(\odot\)})  & 0.12           & 0.78\\\hline
		 Orbital radius (AU)             & 0.029                    & 0.03           \\\hline
		 Planet radius (R\textsubscript{\(\bigoplus\)})  & 0.92     & 12.75      \\\hline

	\end{tabular}
	\end{center}
\end{table}

\subsection{The planet model}\label{sec:model}
As discussed in sections above, the impact of this effect on the overall spectrum has a dependence on $r_{\mathrm{eff}}$ of cloud and also $\theta_{\mathrm{star}}$. The fixed parameters of the model for both the planets are presented in table \ref{tab_model} for the case of TRAPPIST-1e and HD\,189733b. The stellar and planetary radius for HD\,189733b system is taken from \cite{2018A&A...616A...1G} and \cite{2017A&A...602A.107B} respectively. Whereas, those for TRAPPIST-1e system are taken from \cite{Agol_2021}. The planets are considered to have an vertically isothermal atmosphere with the gravity and radius defined at the altitude of model base pressure. The atmosphere is divided in 180 and 126 vertical layers respectively for TRAPPIST-1e and HD\,189733b. The choice of the number of layers is mainly driven by the need to vertically resolve the extent of cloud having the smallest scale height.

\section{Results}\label{sec:Res}
The scattering transparency effect can be produced by a model which can sample the forward scattering phase function (such as the backward Monte Carlo approach of \texttt{scaTran}) and not by models which work by simple geometric ray tracing unless we factor in the non-grey variation of optical depth as mentioned in eq. \ref{eq:dtau} (see section \ref{sec:Comparison}). As discussed in \cite{Robinson_2017} the $MC$ approach considers a Monte Carlo based multiple scattering in the clouds whereas the $geometric$ approach considers the cloud scattering optical depth as extinction along straight rays. First we show the spectrum produced by both the methods i.e. $MC$ approach and $geometric$ approach (without considering the non-grey variation in optical depth). Later, we show that considering a variation in optical depth can also lead to a scattering transparency spectrum.

\begin{figure*}
$\begin{array}{rl}
    \includegraphics[width=0.5\textwidth]{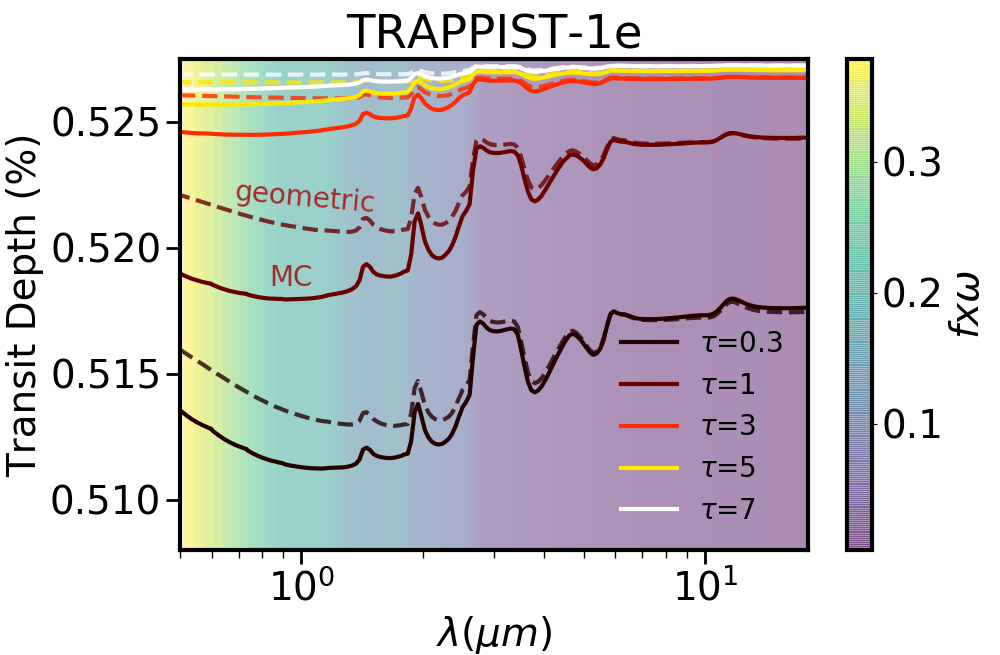}&
    \includegraphics[width=0.5\textwidth]{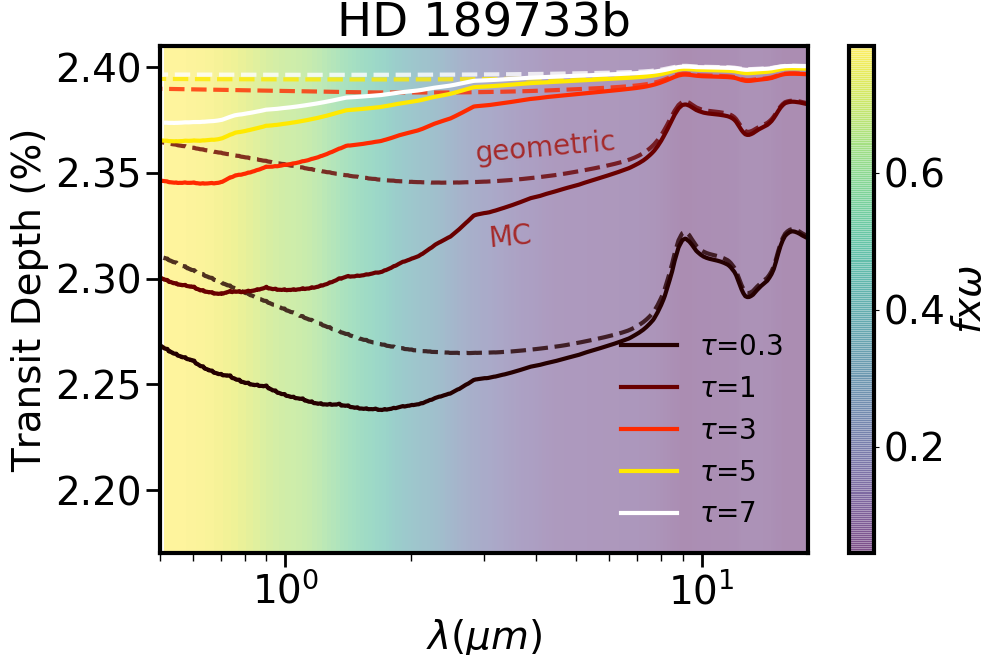}\\
    \multicolumn{2}{c}{\includegraphics[width=0.8\textwidth]{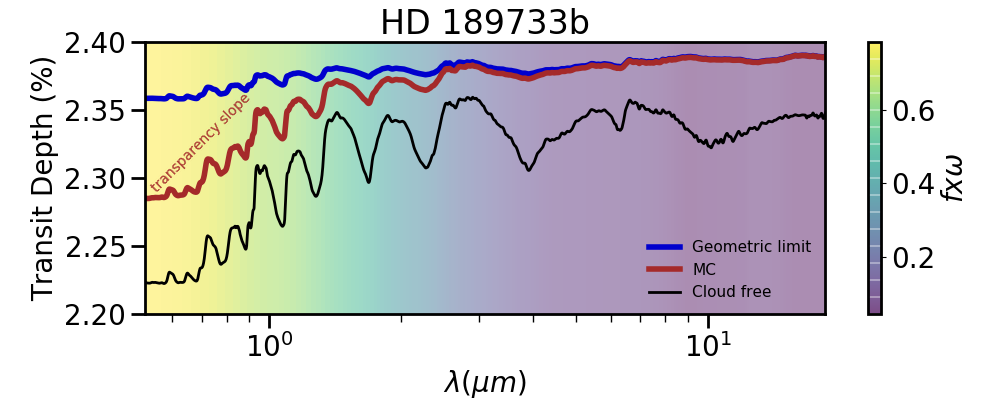}}
\end{array}$

\caption{Scattering transparency effect for various conditions. [Top] Spectral behaviour of this effect for TRAPPIST-1e and HD\,189733b geometry. Results are shown for different slant optical depths $\tau$ for the $geometric$ (dashed lines) and $MC$ (solid lines) cases. [Bottom] The spectrum of HD\,189733b considering absorption by gas as well as clouds, for $\tau=3$. \label{fig:fig5}}
\end{figure*}

The results for various configurations of clouds are presented in figure \ref{fig:fig5}. The top panel shows the cloud spectrum for various cloud scattering optical depths ($\tau$) for the TRAPPIST-1e and HD\,189733b configurations. Here, we do not consider any gas absorption in the medium and show purely the effect of clouds on the spectrum and the spectral features are due to spectral variations of $\omega$ (as in figure \ref{fig:fig2}). However, we do include the Rayleigh scattering from the background gas for both the planets and also the H$_2$ and He collision induced absorptions for HD\,189733b. The cloud scale height is considered to be 1 pressure scale height for both the cases but the cloud base altitude ($Z_{\mathrm{c}}$) is at 100 km ($\sim10$ microbar) and 3000 km ($\sim0.1$ millibar) for TRAPPIST-1e and HD\,189733b respectively. The difference in the $geometric$ and $MC$ cases is clearly observable. The $geometric$ cases mostly produce a flat continuum (except the negative slopes at shorter wavelengths which is due to the Rayleigh scattering from gas) whereas in the $MC$ case the continuum has a positive slope due to the scattering transparency effect. This effect is more pronounced in the hot Jupiter configuration for two reasons: firstly, a larger planet overlap area with the star offers large proportion of scattering particles and secondly, a larger $\theta_{\mathrm{s}}$ for hot Jupiters allows sampling from extended regions of the star (see the variation of $f\times\omega$ with $\theta_{\mathrm{s}}$ in figure \ref{fig:fig3}). Though the scattering transparency effect acts to reduce the transit depth towards shorter wavelengths, at very short wavelengths the Rayleigh scattering by molecules starts to increase the transit depth. For smaller values of $\tau$ (thin clouds), there is a competition between the two processes which can lead to an interesting 'U' shape in the spectrum at around $\sim$1\,$\upmu$m. The color gradient shown in the background of the figures is the theoretically estimated values of $f\times\omega$ which signifies the strength of this effect. One can notice small variations in $f\times\omega$ over the spectrum accompanying the peaks in the spectrum which is due to the variations of $\omega$.

\begin{figure*}
$\begin{array}{rl}
    \includegraphics[width=0.5\textwidth]{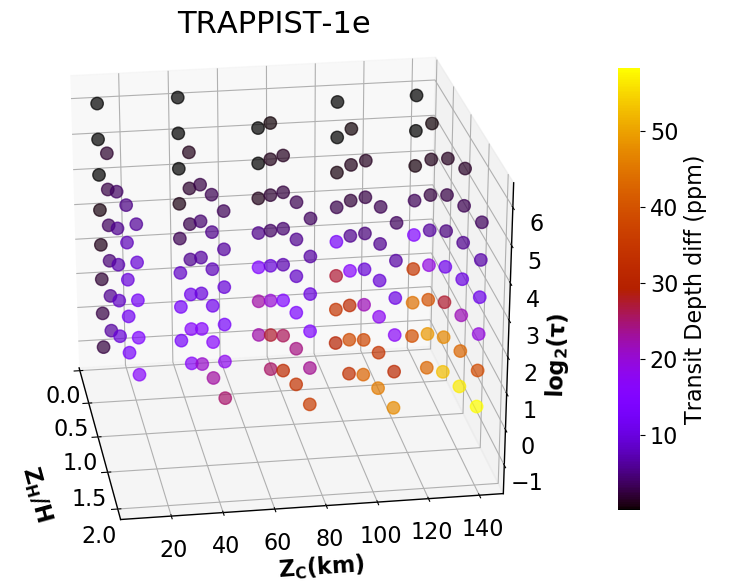}&
    \includegraphics[width=0.5\textwidth]{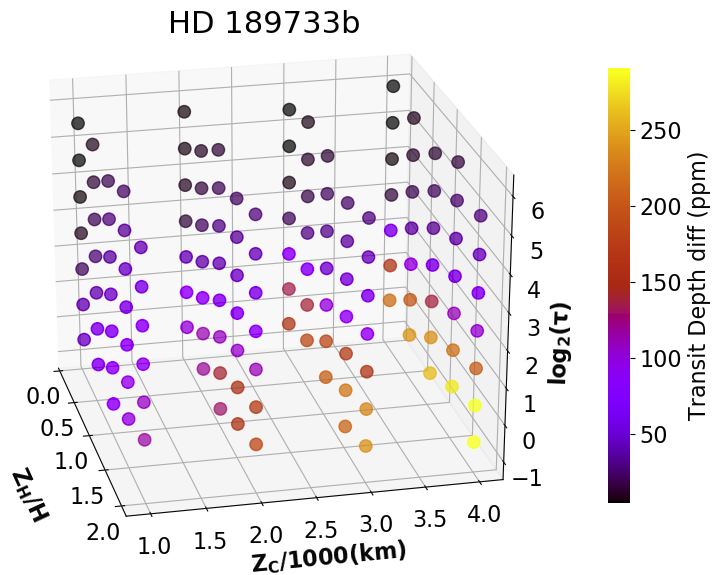}\\
\end{array}$

\caption{Variation in the transit depth (TD) difference due to various cloud parameter are shown for TRAPPIST-1e (TD($\lambda$=5\,$\upmu$m) - TD($\lambda$=0.6\,$\upmu$m)) and HD\,189733b (TD($\lambda$=5\,$\upmu$m) - TD($\lambda$=1.6\,$\upmu$m)).\label{fig:fig6}}
\end{figure*}

\begin{figure*}
\includegraphics[width=0.5\textwidth]{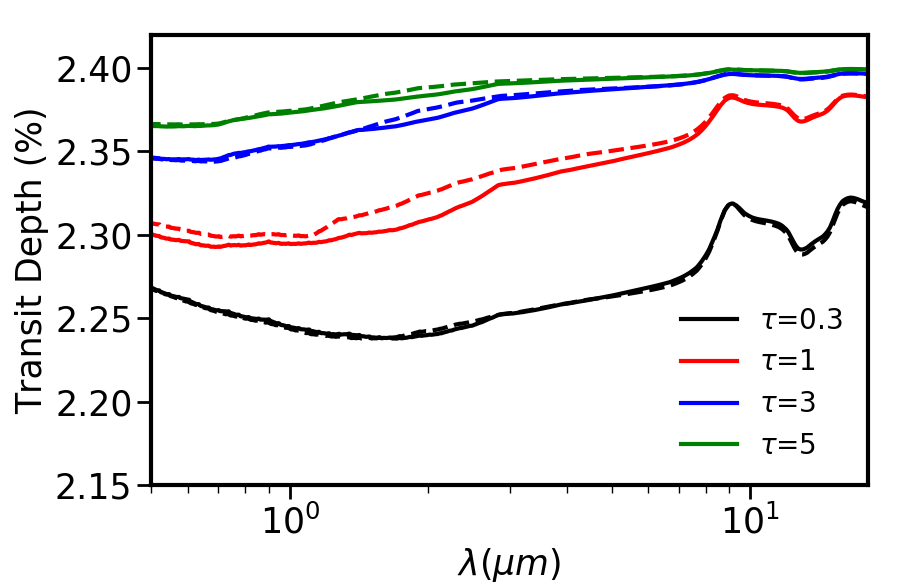}

\caption{A comparison of transit spectra of HD\,189733b simulated with analytical approach (dashed lines) and Monte Carlo approach (solid lines) shown for different values of slant optical depth.\label{fig:fig7}}
\end{figure*}

Finally, we show this effect after including the absorption and scattering by both clouds as well as gas. For this, we include H$_2$O gas absorption opacities \citep{2010JQSRT.111.2139R} for HD\,189733b. Considering the same conditions as in panel (b) for the cloud $\tau$=3 case, we add H$_2$O with a volume mixing ratio of 0.0004 in all the layers. The resultant spectra are shown in the bottom panel of figure \ref{fig:fig5} for $geometric$ and $MC$ cases. The $geometric$ case produces an almost flat spectrum with H$_2$O absorption features whereas the $MC$ case produces a scattering slope continuum with H$_2$O absorption features twice as deep. Towards shorter wavelengths, where scattering transparency is high (high values of $f\times\omega$), the photons can penetrate through the cloud layers when the forward scattering is strong. This allows photons to penetrate to deeper layers where the gas absorption is high. The absorption features of gas (H$_2$O) absorption are seen in the spectrum. For comparison, we also show the spectrum of a $cloud free$ atmosphere. Removing the high altitude cloud leads to deeper gas absorption bands in the transit spectrum as the photons are able to travel to deeper regions of the atmosphere. It is noteworthy that the positive spectral slope in the transit spectrum is caused by the H$_2$O gas opacity. A comparison of the absorption bands in the $MC$ case and the $cloud free$ case reveals how the scattering transparency effect, in $MC$ case, can cause a gradual decrease in the strength of gas absorption from visible/NIR to MIR regions. This variation in the gas absorption band across the spectrum would also be an indicator of scattering transparency of clouds.

We explore parameter space of the cloud profiles for the scattering transparency spectral slopes in figure \ref{fig:fig6}. Considering the NIRSpec (onboard \textit{JWST}) spectral range of 0.6\,$\upmu$m to 5\,$\upmu$m, we consider the transit depth difference at 5\,$\upmu$m and 0.6\,$\upmu$m as a potential metric for predicting the slopes in the \textit{JWST} observations. Other \textit{JWST} instruments (i.e., MIRI) can observe at longer wavelengths and could provide additional constraints on slopes, although noise levels can often be higher in such mid-infrared observations. We explore the effect of clouds using 3 parameters: Cloud base altitude, Cloud scale height and optical depth (slant $\tau$ at cloud base, as in figure \ref{fig:fig1}). Since all of these parameters are poorly-constrained for the exoplanets and given the conditions in which the atmospheres exist, we consider a large range in these parameters. The cloud base is simulated for a range of altitudes. Some of the high altitude cloud cases shown may seem unimaginable (such as clouds at high altitudes having large scale heights) but inspired by a few observations of high altitude haze \citep{2006Icar..183..403M,2021AJ....162...91E} we consider them for the completeness of this study. The cloud scale heights are considered for a large range of $\sim$\,0.1 to $\sim$\,2\,$\times$ the atmospheric scale heights ($H$). It is noteworthy that the cloud vertical profile is considered to be exponential for all the cases and the maximum cloud extent is cutoff at $4\times$ cloud scale height. In our approach to quantify the scattering spectral slope with the difference of two spectral points in the transit spectrum, we do not wish to contaminate it with any kind of absorption slopes. So, to study purely the effect of scattering on cloud parameters, we do not consider any absorption from either the cloud particles or the surrounding gas (similar to $\omega=1$ case of figure \ref{fig:fig4}). The results are shown for TRAPPIST-1e and HD\,189733b transit geometry. Due to strong Rayleigh scattering at lower wavelengths of our HD\,189733b model the lower wavelength point in the Transit Depth difference plots is chosen to be 1.6\,$\upmu$m instead of 0.6\,$\upmu$m. We find that the clouds having, either high cloud base altitudes ($Z_{\mathrm{c}}$) or large scale heights ($Z_{\mathrm{H}}$) are favourable to show stronger effects in the scattering transparency slope. That is because increasing either $Z_{\mathrm{c}}$ or $Z_{\mathrm{H}}$ increases the overlap area of the planetary clouds with the star in a transit geometry leading to an increase in the proportion of the scattered photons which will eventually contribute to the scattering transparency slope in the spectrum. Further, the clouds having low optical depth and low scale heights are also detectable if the cloud is at a higher altitude and has a sufficiently large spatial extent around the terminator. This scenario increases the overlap area of the clouds, because of the regions where stellar rays penetrate below the cloud base altitude (i.e. for $z<Z_{\mathrm{c}}$). For $\tau$ larger than $\sim$10 and $\sim$20 the spectrum remains flat for all the cases for TRAPPIST-1e and HD\,189733b respectively. Within the considered parameter space, we observe that the maximum transit depth difference can reach to no more than 50 ppm and 300 ppm for the two planet cases respectively. Given that the \textit{JWST} error bars on these observations are expected to be $\sim$10-20 ppm (\cite{2016ApJ...817...17G}), these slopes could be observable with \textit{JWST} on rocky worlds and hot Jupiters if such clouds are present.

\subsection{Scattering transparency spectrum with Analytical approach}\label{sec:Comparison}

As discussed earlier, we have shown the results of scattering transparency slopes using a full scale 3D Monte Carlo scattering approach. This approach is accurate but can be time consuming and can take several hours to produce one spectrum. For this reason, it becomes useful to rely on analytical approach in situations requiring quick results. Though analytical approach is not as accurate as an MC calculation it can be a good replacement to MC calculations in various situations. In order to demonstrate the analytical approach we calculate transit spectrum for a non-grey cloud extinction optical depth. The optical depth is varied as per eq. \ref{eq:dtau}, across the spectrum. The scattering transparency spectra obtained with this approach is shown in figure \ref{fig:fig7}  along with the spectra obtained with $MC$ approach. As is seen, the analytical approach produces spectra which closely resemble to those of $MC$ approach and reproduce the scattering transparency slopes. Though the trends in the spectra resemble well, there is a minor difference in the spectra produce by two approaches. This difference is minimum for very small ($\tau<1$) and for very large ($\tau>1$) slant optical depths. For moderate optical depths ($\tau\sim$1), the difference can be as large as $\sim$100 ppm. It happens because the proportion of cloud-scattered photons, reaching the stellar disc, is maximised at moderate optical depths. This leads to an increased difference in the analytical vs. $MC$ approach at moderate optical depths. Though a $MC$ approach to such calculations is an accurate method, we encourage the community to make use of analytical approach for quick estimations.

\section{Discussion}\label{sec:Disc}
The larger and hotter planets such as hot Jupiters may be the best targets to detect the effect of scattering transparency. On smaller terrestrial planets, like TRAPPIST-1 planets, it may be difficult to detect the scattering transparency effect unless the clouds are formed at unusually high altitudes (with respect to terrestrial standards). This is also the reasoning behind our choice of 100 km cloud base altitude for TRAPPIST-1e in figure \ref{fig:fig5}. Here the pressure at the cloud base altitude is $\sim10$ microbars, which is approximately same as for the high altitude carbon dioxide clouds of Mars.

The transit method can only probe the clouds in the terminator region of the planet. So far we have considered uniform clouds on terminators however this condition may not be met in all the targets. The presence of patchy clouds \citep[e.g.][]{2017MNRAS.469.1979M}, for example, can lead to a slope in the spectrum which may not be as high as for the uniform clouds. Similarly, for the planets which may have a difference in cloud properties in east vs. west limb of the planet \citep[see e.g.][]{2019ApJ...887..170P} the scattering slopes may be an aggregate of the two cases of clouds (i.e. on east and west limb).

As we demonstrate here the effect of non-grey Mie scattering, we recommend the future studies of forward scattering from cloud particles to consider this effect for simulations especially for super-micron sized cloud particles. One can start by first estimating the correction term (1-$f\times\omega$) and check its spectral dependence. If the spectral dependence of this term is relatively large, one can either calculate the spectrum using analytical approach or by a full Monte Carlo approach. It may be possible for future studies to improve upon the present work by including scattering phase function of non-spherical cloud grain and also by considering a two term Henyey-Greenstein phase function for a better approximation of forward scattering phase function.

\section{Conclusion}\label{sec:Concl}
Clouds can appear in a wider variety of conditions than considered here. But with the specific conditions considered here, one can say that moderate slant optical depth (1-10), large scale heights and larger $r_{\mathrm{eff}}$ increase our chances of detecting the spectral slopes for a given planet-star system. Many of the assumptions considered here, like a log-normal particle size distribution, width of the distribution, uniformity of clouds on East and West limbs are chosen for brevity but we emphasize that, if all else being equal, larger particles tend to amplify the scattering transparency at shorter wavelengths. This leads to a positive slope in the broadband transit spectrum. This slope exists for strongly forward scattering particles (usually with $r_{\mathrm{eff}}\gtrsim 1$\,$\upmu$m). For particles with $r_{\mathrm{eff}}< 1$\,$\upmu$m the transparency slope starts to reduce significantly as the scattering phase function starts to shrink in the forward direction.

The detection of scattering transparency slopes in the transit spectrum will be an independent indicator of the presence of the clouds having super-micron sized aerosols. In this work, we uncover a viable framework which can be used to predict the effect of scattering transparency over a broad spectral range. Using this framework, it might be possible to infer the important physical properties like $r_{\mathrm{eff}}$ and $\tau$ for the clouds. With \textit{JWST} operational now, we not only can probe the hot Jupiters with better sensitivities but also, for a first time, a large number of cooler worlds. The broadband transit observations of \textit{JWST} for a variety of exoplanets will allow us a chance to validate the predictions of scattering transparency slopes. Clouds having large optical depths or smaller scale heights will eventually lead to a flat spectrum irrespective of cloud particle size. However, given the variety of exoplanet atmospheres which will be probed by \textit{JWST} with its large wavelength coverage, we envisage the detection of the scattering transparency effect on several targets.



\section*{Acknowledgement}
We acknowledge many comments and suggestions from the reviewer which have helped us to improve the quality of the manuscript. BJ acknowledges the initial guidance and discussions with Nikku Madhusudhan which have led to this investigation.

\section*{Data Availability}
The data underlying this article will be shared on reasonable request to the corresponding author.



\bibliographystyle{mnras}
\bibliography{bibliography} 




\appendix


\bsp	
\label{lastpage}
\end{document}